\documentclass[aps,prd,twocolumn,tightelines,nofootinbib]{revtex4}
\usepackage{graphicx}
\usepackage{epsfig}
\usepackage{bm}
\usepackage{latexsym,amssymb,amsmath,amsfonts,amssymb,txfonts,pxfonts,wasysym,float}
\usepackage{mathrsfs}
\usepackage{color}
\usepackage{lettrine}
\usepackage{lipsum}
\usepackage{enumitem}

\newcommand{\postscript}[2]{\setlength{\epsfxsize}{#2\hsize}
   \centerline{\epsfbox{#1}}}

\usepackage[usenames,dvipsnames]{xcolor}
\definecolor{orange}{cmyk}{0,0.5,1,0}
\definecolor{rossoCP3}{cmyk}{0,.88,.77,.40}
\definecolor{graa}{rgb}{0.8,0.8,0.8}
\definecolor{blaa}{rgb}{0.2,0.2,0.6}

\begin{document}

\title{\color{rossoCP3}
A physicist view of COVID-19 airborne infection through convective
airflow in indoor spaces}

\author{Luis A. Anchordoqui}
\affiliation{Physics Department, Herbert H. Lehman College and Graduate School, The City University of New York\\ 250 Bedford Park Boulevard West, Bronx, New York 10468-1589, USA}

\author{Eugene M. Chudnovsky}

\affiliation{Physics Department, Herbert H. Lehman College and Graduate School, The City University of New York\\ 250 Bedford Park Boulevard West, Bronx, New York 10468-1589, USA}

\date{March 2020} 

\begin{abstract}
  \noindent {\it General Idea:} Naturally produced droplets from
  humans (such as those produced by breathing, talking, sneezing, and
  coughing) include several types of cells (e.g., epithelial cells and
  cells of the immune system), physiological electrolytes contained in
  mucous and saliva (e.g. Na$^+$, K$^+$, Cl$^-$), as well as,
  potentially, several infectious agents (e.g. bacteria, fungi, and
  viruses). In response to the novel coronavirus SARS-CoV-2 epidemic,
  which has become a major public health issue worldwide, we provide a
  concise overview of airborne germ transmission as seen from a
  physics perspective. We also study whether coronavirus aerosols can
  travel far from the immediate neighborhood and get airborne with the
  convective currents developed within confined spaces.\\  {\it
    Methodology:} Methods of fluid dynamics are utilized to analyze the
  behavior of various-size airborne droplets containing the virus. \\
  {\it Study
    Findings:} We show that existing vortices in the air can make a
  location far away from the source of the virus be more dangerous
  than a nearby (e.g., 6 feet away) location.\\ {\it Practical
    Implications}: Our study reveals that it seems
  reasonable to adopt additional infection-control measures to the
  recommended 6~feet social distancing. We provide a recommendation
  that could help to slow down the spread of the virus.
 \end{abstract}
\maketitle

\section{Introduction}

The recent outbreak of the respiratory disease identified as COVID-19
is
caused by the severe acute respiratory syndrome coronavirus 2,
shortened to SARS-CoV- 2~\cite{Huang,Zhou,Zhu}. The outbreak,
first reported in December 2019, 
has rapidly evolved into a global pandemic. Indeed, COVID-19 is spreading across the globe with a speed and strength that laid
bare the limits of our understanding of the transmission pathways and
the associated factors that are key to the spread of such
diseases. In particular, the virus can spread from seemingly healthy
carriers or people who had not yet developed symptoms~\cite{Rothe}. Overall, this
has transformed the face of healthcare around the world.

To understand and prevent the
spread of a virus like SARS-CoV-2, it is important to estimate the probability of
airborne transmission as aerosolization with particles potentially
containing the virus. There have been  reports favoring the
possibility of creating coronavirus aerosols~\cite{Doremalen}. Thus
far no aerosolized coronavirus particles have been found in hospital searches
within the most public areas, but evidence has been detected in rooms of SARS-CoV-2 patients~\cite{Ong,Santarpia,Liu:20,Guo}. In this paper we
provide an overview on the possible threat of SARS-CoV-2
airborne infection from a physics point of view, focusing attention
on the effect of convection currents in indoor spaces.
 
The layout of the paper is as follows. In Sec.~\ref{sec:2} we first
provide a concise discussion of the motion of the virus in suspended
aerosols. After that, using the SimScale program~\cite{simscale} we study
the convective airflow in a meeting room and office space. We show
that existing vortices in the air can make a location far away from
the source of the virus be more dangerous than a nearby (e.g., 6 feet away) location.  In Sec.~\ref{sec:3} we present our conclusions. Throughout we adopt the convention of the World Health
Organization to nickname  
particles that are $\agt 5~\mu$m diameter as droplets and those $\alt
5~\mu$m as aerosols or droplet nuclei~\cite{Shiu}.

\section{Modeling the effect of convection currents in the
  transmission of SARS-COV-2}
\label{sec:2}

In the presence of air resistance, compact heavy objects fall to the ground quickly, while light objects exhibit
Brownian motion and follow the pattern of turbulent convection of the
air. For aerosol particles containing the virus, the boundary between
these two behaviors depends on the size of the particle. We begin with
a simple question: how long does a virus float in the air under the influence of gravity?  To answer this query we
model the virus as a sphere of radius $r \sim 90~{\rm nm}$ and mass 
$m \sim 2.5 \times 10^{-19}~{\rm kg}$~\cite{Woo}, and we assume that this spherical particle is suspended
in a viscous fluid (the air) feeling the Earth's gravitational
field. Herein, gravity tends to make the particles settle, while diffusion
and convection act to homogenize them, driving them into regions of
smaller concentration. On the one hand, the convection mechanism provides particle
macro-mixing within the fluid through the tendency of hotter and
consequently less dense material to rise, and colder, denser material
to sink under the influence of gravity. On the other hand, the diffusion mechanism acts on the scale of an individual particle (micro-mixing)
slowly and randomly moving through the media.

Under the action of
gravity, the virus acquires a downward terminal speed that follows from Stokes
 law and is given by
\begin{equation}
  v_{\rm down} = \mu m g \,,
\label{eq1}
\end{equation}
 where $g \simeq 9.8~{\rm m/s^2}$ is the acceleration due to gravity and
 \begin{equation}
  \mu = \frac{1}{6 \pi \eta r} \,,
  \label{eq2}
\end{equation}
is the virus mobility in the fluid,
and where $\eta = 1.8 \times 10^{-5}~{\rm kg/(m s)}$ is the dynamic
viscosity of air~\cite{Einarsson}. Substituting (\ref{eq2}) into (\ref{eq1}) we find
that the downward terminal speed of the virus in dry air is indeed negligible,
$v_{\rm down} \sim 8 \times 10^{-8}~{\rm m/s}$. It is therefore clear
that gravity plays no role in the motion of an isolated virus through
the air. Rather it follows a convection pattern in a manner similar to
how smelly substances move through the air.  The survival
probability of the virus in the dry air is then given by the
likelihood of survival outside its natural environment. The half-life of SARS-CoV-2 in aerosols has been found to be about 1.1 hours~\cite{Doremalen}.

\begin{table}
\caption{
Evaporation time of water droplets. \label{tabla}}
  \begin{tabular}{cc}
    \hline
    \hline
    ~~~~Droplet diameter ($\mu$m)~~~~ & ~~~~Evaporation time (s)~~~~ \\
   \hline 
$2000$ & $660\phantom{.0}$ \\
$1000$ & $165\phantom{.0}$ \\
    $\phantom{0}500$ & $\phantom{0}41\phantom{.0}$ \\
    $\phantom{0}200$ & $\phantom{00}6.6$ \\
    $\phantom{0}100$ & $\phantom{00}1.7$ \\
    $\phantom{00}50$ & $\phantom{00}0.4$ \\
    \hline
    \hline
     \end{tabular}
   \end{table}

\begin{figure*}[tb]
\begin{minipage}[t]{0.49\textwidth}
\postscript{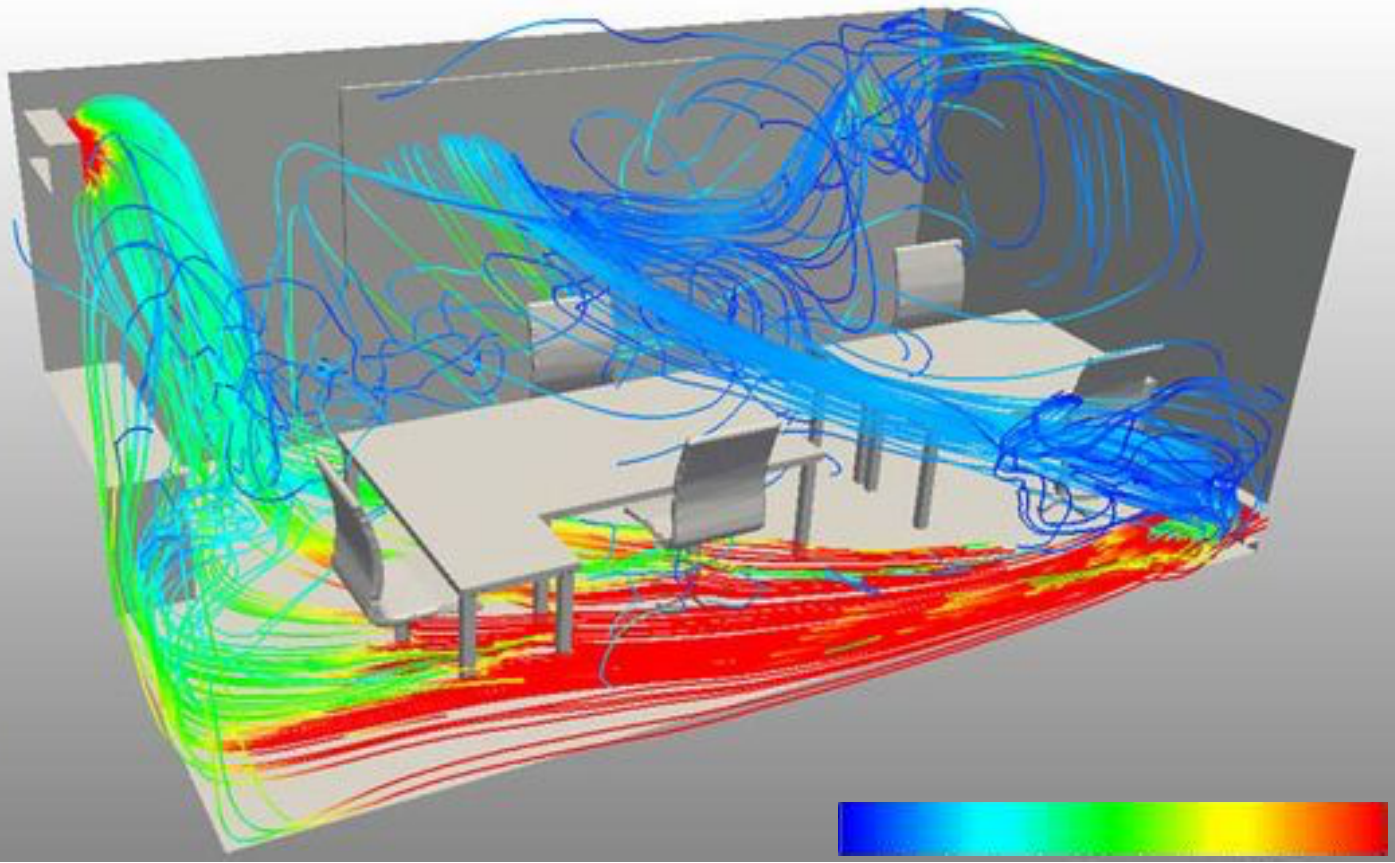}{0.9}
\end{minipage}
\begin{minipage}[t]{0.49\textwidth}
\postscript{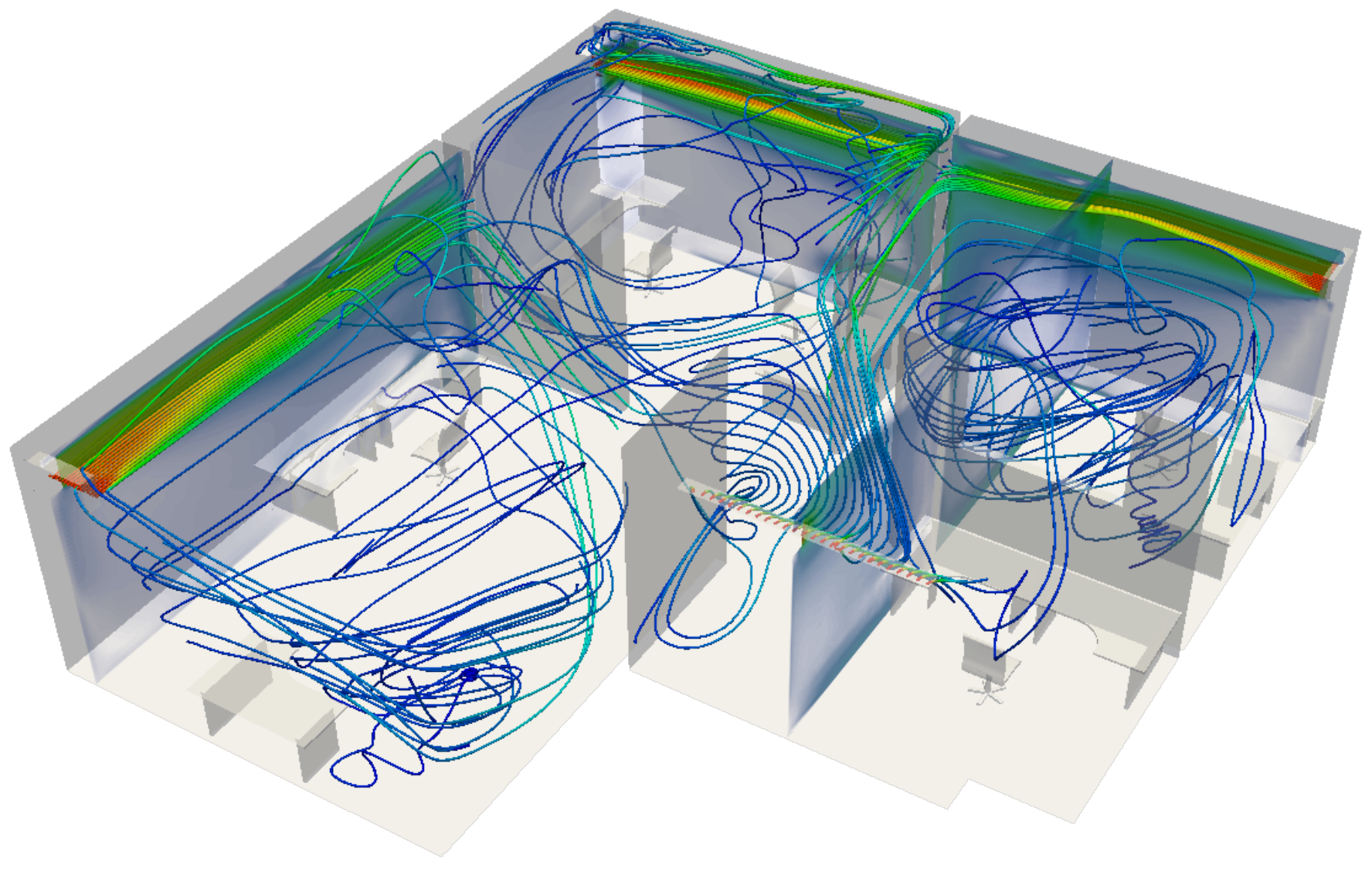}{0.9}
\end{minipage}
\caption{Visual representation of airflow streamlines in a meeting
  room (left) and office
  space (right) colored to velocity magnitude from low (blue) to high
  (red). The convection pattern in the meeting room demonstrates how the infection
can be persistently carried by the airflow between two chairs
separated by 6 feet. The convection pattern in the office space
illustrates how the infection can be taken by the airflow from one
cubicle to the other. Simulation by SimScale~\cite{simscale}. \label{fig:1}}
\end{figure*}
We have seen that the coronavirus can go airborne staying suspended
in the air. However, the virus is transmitted through respiratory
droplets and droplet nuclei produced mostly while sneezing and coughing. Then
 to ascertain whether airborne transmissible SARS-CoV-2 can survive and
stay infectious in aerosols we must double-check that the
 falling time of a droplet or a droplet nuclei from a height of about two meters is larger than its
evaporation time scale. To this end, we assume that the drops are
also spherical and hence the mass can be simply estimated as
\begin{equation}
  m = \frac{4}{3} \pi r^3 \ \rho \,,
\end{equation}
where $\rho = 997~{\rm kg/m^3}$ is the density of water and $r$ the
droplet/aerosol radius. For large droplets whose diameters $\agt 1000~\mu{\rm m}$,
the effect of air resistance is negligible and so the falling time 
can be directly estimated using Newton's equations for gravitational
settling. For smaller droplets whose diameters $ <100~\mu{\rm m}$, the
falling times must instead be determined using the downward terminal speed
given in (\ref{eq1}) to account for the air resistance upon the
falling droplets. It is now an instructive and straightforward exercise to
show that the time for falling 2~m in saturated air is 0.6~s for
droplets with $r> 500~\mu{\rm m}$, 6.0~s for those of
$r \sim 50~\mu{\rm m}$, 600~s (about 10 minutes) for those of
$r\sim 5~\mu{\rm m}$, and 60,000~s (about 16.6 hours) for those of
$r \sim 0.5~\mu{\rm m}$. The droplet evaporation time scale, as
computed by Wells using droplet evaporation data
collected by Whytlaw-Gray and Patterson, is shown in Table~\ref{tabla}~\cite{Wells}. The
assumption of pure water droplets in unsaturated air at $18^\circ~{\rm
  C}$ was used for the
evaporation calculations, such that the theoretical droplets are
capable of complete evaporation. By direct comparison of the droplet
evaporation and 
falling times we can conclude that somewhere between 100 and
200~$\mu$m lies the droplet size (i.e. the diameter) which identifies droplets of
mouth spray that reach the ground within the life of the droplet as
against droplets that evaporate and remain in the air as {\it
  droplet-nuclei} with attached SARS-CoV-2 infection. Several investigations have been carried out to continue improving the precision
of Wells analyses and to study the various external environmental
(such as temperature and humidity) factors that
may alter his estimates; see e.g.~\cite{Liu}. 
  
The sizes of the droplets and droplet-nuclei
produced by sneezing and coughing 
were studied by the microscopic measurement of  12,000 droplet stain-marks found on slides exposed directly to
mouth-spray, and of 21,000 stain-containing droplet-nuclei recovered
from the air on to oiled slides exposed in the slit sampler~\cite{Duguid}. From this
data sample it was found that the original diameters of respiratory
droplets ranged from 1 to 2000~$\mu$m and that 95\% were between 2
and 100~$\mu$m and that the most common were between 4 and
8~$\mu$m. Similar results were reported in~\cite{Gralton}. This suggest that, in principle, droplet-spray could drive
direct airborne infection of SARS-CoV-2. The transmission of the
COVID-19 disease, however, still depends on the infectious virus load
carried by the droplets, which must be determined
experimentally. The number of virions needed for infection is yet
unknown, but we can use other viral transmission (e.g., influenza~\cite{Yang,Yezli,Nikitin}) for
a template. The spread of a sneeze in the
air has been studied by ultrafast imaging at
MIT~\cite{Bourouiba,Scharfman,Bourouiba2}. It was found that even the largest droplets from a sneeze can float in the air for up to 10 minutes, which allows them to reach the far end of a large room. This points towards convection in the air being the primary mechanism of the spread of the infection.

From the physics point of view, we
cannot find a good justification for a stationary 6-feet separation in
a situation when people spend long time together in a room. Small droplets or aerosols
containing the virus move in the air via convection. The convection
pattern in a room can be very complex; see Fig.~\ref{fig:1}. It
depends on the location of air conditioners, radiators, windows, and
all items in the room, as well as on people producing vortices by
moving around. The existing vortices in the air can make a location
far away from the source of the virus more dangerous than the location
6 feet away. This applies to meeting rooms, office spaces,
supermarkets, department stores, etc. The airflow pattern should be
studied for all such facilities to avoid the spread of infection to
large distances from a single infected person. The safest rooms must
be those equipped with the air sucking ventilator at the top, like
hospital surgery rooms~\cite{Yu}.

By all means, re-configuring the ventilation of public and private facilities cannot be done within the timescale of the pandemic. The question is what to do now if we want to slow its pace. The answer is very simple. People must be required to wear face masks in public spaces to prevent the virus from becoming airborne in the first place.

\section{Conclusions}
\label{sec:3}

Airborne SARS-CoV-2 virus spreads from infected individuals and
accumulates in confined spaces where it can linger in the air in the
aerosol form for hours. The inhaled virus load depends on the virus
concentration in the air and the time of exposure. Concentration can
vary from one spot to another. It depends on the location of the
spreaders and the pattern of the airflow. The latter is determined by
many factors, such as the location of doors and windows, ventilators,
heaters, movement of people, etc.  Common central air conditioning
system that is cooling the indoor air but is not exchanging it with
the outside air and is not filtering the virus helps to spread it
across the airconditioned space. It can easily take the virus to
distant locations and make it accumulate in the least expected
places. Computer studies of the airflow in public spaces are important
for mitigating this problem. In the long run business and educational
facilities should consider redesigning ventilation and air
conditioning systems to effectively reduce concentration of the
aerosol virus in the air.\\

\noindent{\bf Funding/Support:} The theoretical and computational techniques and resources used in
this research were supported by the U.S. National Science Foundation,
NSF Grant PHY-1620661 (L.A.A.), and the U.S. Department of Energy,
Office of Science, DOE Grant DE- FG02-93ER45487 (E.C.).\\

\noindent {\bf Role of the Funder/Sponsor:} The sponsors had no role in
the preparation, review or approval of the manuscript and decision to
submit the manuscript for publication. Any opinions,
findings, and conclusions or recommendations expressed in this
article are those of the authors and do not necessarily reflect the
views of the NSF or DOE.\\

\noindent {\bf Conflict of Interest Disclosures:} None.


\begin{thebibliography}{99}


  
\bibitem{Huang} C. Huang, Y. Wang, X. Li, L. Ren, J. Zhao, Y. Hu, L. Zhang, G. Fan,
J. Xu, X. Gu, Z. Cheng, T. Yu, J. Xia, Y. Wei, W. Wu, X. Xie, W. Yin,
H. Li, M. Liu, Y. Xiao, H. Gao, L. Guo, J. Xie, G. Wang, R. Jiang,
Z. Gao, Q. Jin, J. Wang, and B. Cao, {\color{rossoCP3} Clinical features of patients infected with 2019 novel coronavirus in Wuhan, China},
Lancet {\bf 395}, 497 (2020)
doi:10.1016/S0140-6736(20)30183-5




\bibitem{Zhou} P. Zhou, X. Yang, X. Wang, B. Hu, L. Zhang, W. Zhang, H. Si, Y. Zhu,
B. Li, C. Huang, H. Chen, J. Chen, Y. Luo, H. Guo, R. Jiang, M. Liu,
Y. Chen, X. Shen, X. Wang, X. Zheng, K. Zhao, Q. Chen, F. Deng,
L. Liu, B. Yan, F. Zhan, Y. Wang, G. Xiao, and Z. Shi,
 {\color{rossoCP3} A pneumonia outbreak associated with a new
   coronavirus of probable bat origin}, Nature {\bf 579}, 270 (2020)
doi:10.1038/s41586-020-2012-7
 
\bibitem{Zhu}
N. Zhu, D. Zhang, W. Wang, X. Li,  B. Yang, J. Song, X. Zhao,  B. Huang, W. Shi, R. Lu, P. Niu, F. Zhan, X. Ma, D. Wang, W. Xu,  G. Wu, G. F. Gao, and W. Tan, 
 {\color{rossoCP3} A novel coronavirus from patients with pneumonia in China, 2019},
N. Engl. J. Med. {\bf 382}, 727 (2020).
doi:10.1056/NEJMoa2001017

\bibitem{Rothe}
C. Rothe, M. Schunk, P. Sothmann, G. Bretzel, G. Froeschl, C. Wallrauch, T. Zimmer, V. Thiel, C. Janke, W. Guggemos, M. Seilmaier, C. Drosten, P. Vollmar, K. Zwirglmaier, S. Zange, R. W\"olfel, M. Hoelscher
 {\color{rossoCP3}
Transmission of 2019-nCoV Infection from an Asymptomatic Contact in Germany},
N. Engl. J. Med. {\bf 382}, 970 (2020).
doi:10.1056/NEJMc2001468


\bibitem{Doremalen}
  N. van Doremalen, T. Bushmaker, D. H. Morris, M. G. Holbrook,
  A. Gamble, B. N. Williamson, A. Tamin, J. L. Harcourt,
  N. J. Thornburg, S. I. Gerber, J. O. Lloyd-Smith, E. de Wit, and
  V. J. Munster,
 {\color{rossoCP3} Aerosol and surface stability of SARS-CoV-2 as compared with SARS-CoV-1},
 N. Engl. J. Med. {\bf 382}, 1564 (2020). 
doi:10.1056/NEJMc2004973

\bibitem{Ong}
  S. W. X. Ong, Y. K. Tan, P. Y. Chia, T. H. Lee, O. T. Ng,
  M. S. Y. Wong, and K. Marimuthu, 
{\color{rossoCP3} Air, surface environmental, and personal protective equipment contamination by severe acute respiratory syndrome coronavirus 2 (SARS-CoV-2) from a symptomatic patient},
JAMA {\bf 323},  1610 (2020).
doi:10.1001/jama.2020.3227

\bibitem{Santarpia}
  J. L. Santarpia, D. N. Rivera, V. Herrera, M. J. Morwitzer, H. Creager, G. W. Santarpia, K. K. Crown, D. M. Brett-Major, E.
Schnaubelt, M. J. Broadhurst, J. V. Lawler, St. P. Reid, and
J. J. Lowe,
  {\color{rossoCP3}  Transmission potential of SARS-CoV-2 in viral
    shedding observed at the University of Nebraska Medical Center},
medRxiv preprint
doi:10.1101/2020.03.23.20039446

\bibitem{Liu:20} Y. Liu, Z. Ning, Y. Chen, M. Guo, Y. Liu, N. K. Gali, L. Sun, Y. Duan, J. Cai, D. Westerdahl, X. Liu, K. Xu, K.-f. Ho, H. Kan, Q. Fu, and K. Lan,
{\color{rossoCP3} Aerodynamic analysis of SARS-CoV-2 in two Wuhan hospitals},
Nature (2020)
doi:10.1038/s41586-020-2271-3



\bibitem{Guo}
Z.-D. Guo, Z.-Y. Wang, S.-F. Zhang, X. Li, L. Li, C. Li, Y. Cui, R.-B. Fu, Y.-Z. Dong, X.-Y. Chi, M.-Y. Zhang, K. Liu, C. Cao, B. Liu, K. Zhang, Y.-W. Gao, B. Lu, and W. Chen,
{\color{rossoCP3} Aerosol and surface distribution of severe acute respiratory syndrome coronavirus 2 in hospital wards, Wuhan, China, 2020}, 
Emerg. Infect. Dis. {\bf 26}, 1583 (2020)
doi:10.3201/eid2607.200885

\bibitem{simscale}
  {\tt https://www.simscale.com}


\bibitem{Shiu}
E. Y. C. Shiu, N. H. L. Leung, and B. J. Cowling,
{\color{rossoCP3} Controversy aorund airborne versus droplet transmission of respiratory viruses: implications for infection prevention},
Curr. Opin. Infect. Dis. {\bf 32}, 372 (2019)
doi:10.1097/QCO.0000000000000563



\bibitem{Woo}
  P. C. Y. Woo, Y. Huang, S. K. P. Lau, and K. Y. Yuen,
 {\color{rossoCP3}  Coronavirus genomics and bioinformatics analysis},
Viruses {\bf 2}, 1804 (2010).
doi:10.3390/v2081803

\bibitem{Einarsson}
  J. Einarsson and  B. Mehlig,
 {\color{rossoCP3} Spherical particle sedimenting in weakly
   viscoelastic shear flow},
 Phys. Rev. Fluids {\bf 2}, 063301 (2017).
 doi:10.1103/PhysRevFluids.2.063301

\bibitem{Wells}
  W. F. Wells,
 {\color{rossoCP3} On air-borne infection study II: droplets and
   droplet nuclei},
 Am. J. Hyg.  {\bf 20}, 611 (1934).




\bibitem{Liu}
  F. Liu, H. Qian, X. Zheng，J. Song, G. Cao, and Z. Liu,
 {\color{rossoCP3} Evaporation and dispersion of exhaled droplets in stratified environment},
 IOP Conf. Ser. Mater. Sci. Eng. {\bf 609} 042059 (2019).
 doi:10.1088/1757-899X/609/4/042059


 
\bibitem{Duguid}
  J. P. Duguid,
 {\color{rossoCP3}  The size and the duration of air-carriage of respiratory droplets and droplet-nuclei},
 J. Hyg. (Lond.) {\bf 44}, 471 (1946).
doi:10.1017/s0022172400019288


\bibitem{Gralton}
  J. Gralton, E. Tovey, M. L. McLaws, and W. D. Rawlinson, 
 {\color{rossoCP3}  The role of particle size in aerosolised pathogen
   transmission: A review},
J. Infect. {\bf 62}, 1 (2011)
 doi:10.1016/j.jinf.2010.11.010

\bibitem{Yang}
  W. Yang, S. Elankumaran, and L. C. Marr,
{\color{rossoCP3}  Concentrations and size distributions of airborne influenza A viruses measured indoors at a health centre, a day-care centre and on aeroplanes}
J. R. Soc. Interface {\bf 8}, 1176 (2011)
doi:10.1098/rsif.2010.0686

\bibitem{Yezli}
  S. Yezli and J. A. Otter,
  {\color{rossoCP3} Minimum infective dose of the major human respiratory and enteric viruses transmitted through food
and the environment},
Food Environ. Virol. {\bf 3}, 1 (2011)
doi:10.1007/s12560-011-9056-7

\bibitem{Nikitin}
  N. Nikitin, E. Petrova, E. Trifonova, and O. Karpova,
 {\color{rossoCP3} Influenza virus aerosols in the air and their
   infectiousness},
 Adv. Virol. {\bf 2014}, 859090 (2014)
doi: 10.1155/2014/859090



\bibitem{Bourouiba}
  L. Bourouiba, E. Dehandschoewercker, and  J. W. M. J. Bush,
 {\color{rossoCP3} Violent expiratory events: on coughing and sneezing},
Fluid Mech. {\bf 745}, 537 (2014).
doi:10.1017/jfm.2014.88

\bibitem{Scharfman}
B. E. Scharfman, A. H. Techet,  J. W. M. Bush, and  L. Bourouiba,
  {\color{rossoCP3}  Visualization of sneeze ejecta: steps of fluid fragmentation leading to respiratory droplets},
L. Exp. Fluids {\bf 57}, 24 (2016).
doi:10.1007/s00348-015-2078-4

\bibitem{Bourouiba2}
 L. Bourouiba,
 {\color{rossoCP3}  Turbulent gas clouds and respiratory pathogen emissions:
Potential implications for reducing transmission of COVID-19},
JAMA {\bf 323}, 1837 (2020).
doi:10.1001/jama.2020.4756



\bibitem{Yu}
 H. C. Yu, K. W. Mui, L. T. Wong, and H. S. Chu,
  {\color{rossoCP3} Ventilation of general hospital wards for mitigating infection risks of three kinds of viruses including Middle East respiratory syndrome coronavirus}
Indoor and Built Environment {\bf 26}, 514 (2016).
doi.org/10.1177/1420326X16631596
  
\end{thebibliography}
\end{document}